\documentclass[aps,prb,twocolumn,groupedaddress]{revtex4-1}

\usepackage{graphics}
\newcommand{\erti}{Er$_2$Ti$_2$O$_7$}
\newcommand{\mub}{$\mu_{\rm B}$}

\begin{document}

\title{Field evolution of the magnetic structures in
Er$_2$Ti$_2$O$_7$ through the critical point.}
\author{H.B. Cao}
\author{I. Mirebeau}
\email[]{isabelle.mirebeau@cea.fr}
\author{A. Gukasov}
\affiliation{CEA, Centre de Saclay, DSM/IRAMIS/Laboratoire L\'eon Brillouin,
91191 Gif-sur-Yvette, France}
\author{P. Bonville}
\affiliation{CEA, Centre de Saclay, DSM/IRAMIS/Service de Physique de l'Etat
Condens\'e, 91191 Gif-sur-Yvette, France}
\author{C. Decorse}
\affiliation{Laboratoire de Physico-Chimie de l'Etat Solide, ICMMO,
Universit\'e Paris-Sud, 91405 Orsay, France}

\date{\today}

\begin{abstract}
We have measured neutron diffraction patterns in a single crystal sample
of the pyrochlore compound Er$_2$Ti$_2$O$_7$ in the antiferromagnetic
phase (T=0.3\,K), as a function of the magnetic field, up to 6\,T, applied along the [110] direction. We determine all the characteristics of the magnetic structure throughout the
quantum critical point at $H_c$=2\,T. As a main result, all Er moments align along the
field at $H_c$ and their values reach a minimum.
Using a four-sublattice self-consistent calculation, we show that the
evolution of the magnetic structure and the value of the critical field are rather well reproduced using the same anisotropic exchange tensor as that accounting for the local paramagnetic susceptibility. In contrast, an isotropic exchange tensor does not match the moment variations through the critical point. The model also
accounts semi-quantitatively  for other experimental data previously measured, such as the field dependence
of the heat capacity, energy of the dispersionless inelastic modes and
transition temperature.

\end{abstract}


\maketitle
\section{Introduction}
Geometrical magnetic frustration allows one to obtain materials with tunable
properties. It yields a large landscape of possible magnetic ground states,
due to the inability of the system to choose a unique spin configuration
which minimizes the energy. Rare earth pyrochlores R$_2$Ti$_2$O$_7$, where the
R magnetic moments reside on the summits of corner sharing tetrahedra, are
model systems to study such effects. Here, geometrical frustration does not
arise from the competition of magnetic interactions, but rather emerges in
the context of a highly symmetrical structure, from the subtle interplay of
three main energy terms: the  single ion crystal field anisotropy, the
exchange interaction and the magnetic dipolar coupling. In R$_2$Ti$_2$O$_7$,
the trigonal symmetry of the crystal field at the R site comes from the oxygen
environment. At low temperature, this yields two generic behaviors, Ising-like
(Ho, Dy, Tb) or XY-like (Er, Yb), depending on whether the $<$111$>$ axes are
easy or hard anisotropy axes for the magnetic moments. The final selection of a
magnetic state within the ground state manifold is determined by the nature,
lengthscale and sign of the magnetic interaction, by perturbation energy
terms, or by $"$order by disorder$"$ processes \cite{Villain80,Chalker92}.

The spin ice compounds Ho$_2$Ti$_2$O$_7$ and Dy$_2$Ti$_2$O$_7$, combining a
strong Ising anisotropy with an effective ferromagnetic exchange, show exotic
short range orders with macroscopic entropy \cite{Bramwell012} and peculiar
excitations, where the R dipolar moments fractionalize into magnetic monopoles
\cite{Castelnovo08}. Here we focus on Er$_2$Ti$_2$O$_7$, a compound with
``reversed'' behavior, namely with planar XY anisotropy and antiferromagnetic
(AF) interactions, and showing magnetic order below 1.2\,K \cite{Blote}.

Er$_2$Ti$_2$O$_7$ was proposed to realize a model type XY antiferromagnet, for
which theory predicts a fluctuation induced symmetry breaking, leading to
magnetic long range ordering \cite{Champion03}. Indeed, for the crystal field ground
Kramers doublet of the Er$^{3+}$ ion, the threefold symmetry [111] axis is a hard magnetic
axis, the easy plane being the local (111) plane. Below $T_{\rm N}$=1.2\,K, the
antiferromagnetic structure has a {\bf k}=0 propagation vector and it is defined
by the basis vectors $\psi_2$, which transform according to the irreducible representation
$\Gamma_5$, following the Kovalev notations used in Refs.\onlinecite{Champion03} and
\onlinecite{Poole07}.
It consists of six equally populated domains, as shown by spherical neutron
polarimetry \cite{Poole07}. The selection of this particular state among the
possible basis states of the {\bf k}=0 manifold is still subject to
discussion. Surprisingly, it is the only non-coplanar structure among all
others, whereas order by disorder processes usually select coplanar or
collinear orders \cite{Chalker10}. It also differs from the the so-called
Palmer-Chalker state \cite{Palmer00}, predicted to be the ground state in the
presence of AF
isotropic exchange and dipolar interactions. An energetic selection of the
ordered states was recently proposed \cite{Clarty09}, arising from sixth order
terms in the crystal field Hamiltonian, together with possible anisotropic
exchange, dipolar and Dzyaloshinskii interactions. An anisotropic molecular
field tensor with antiferromagnetic components, reinforcing the crystal field
planar anisotropy, has indeed been shown to be present in Er$_2$Ti$_2$O$_7$ using
polarized neutron diffraction, by
measuring the local susceptibility in the paramagnetic phase \cite{Cao09}.

The magnetic structure near $T$=0 has been shown to coexist with spin
fluctuations by $\mu$SR spectroscopy \cite{Lago05}: a non-vanishing spin
dynamics washes out the precession signal usually observed by muons in ordered
magnets. This is confirmed by the presence of soft collective excitations
probed by inelastic neutron scattering \cite{Ruff08}. In this latter work,
it is shown that application of a magnetic field (along [110])
decreases the N\'eel temperature, resulting in a zero temperature phase
transition at a critical field $H_c \simeq 1.8$\,T. The magnetic ordered
state is suggested to transform into some kind of spin liquid state above the
critical field, through a second order quantum phase
transition driven by spin fluctuations \cite{Ruff08, Clarty09}.
The field evolution of the magnetic
structure was qualitatively understood as a smooth deformation from the zero
field $\psi_2$ configuration to a field configuration above $H_c$ where the
moments are aligned with or close to the field direction. In these studies,
however, the spin configurations could not be characterised in detail,
considering the limited information provided by the cold neutron measurements.

In this work, we present detailed in-field neutron diffraction experiments in the
AF phase of \erti\ using hot neutrons. This allows us to follow the evolution of the
ground state induced by a magnetic field applied along [110], through the quantum
critical point and up to a field of 6\,T, and to clarify both questions of the
domain structure and of the microscopic local
spin structure within a tetrahedron. We show that the Er moment magnitude presents
a minimum around $H_c$, which can be considered as the spin-flip field for the [110]
direction. Up to 6\,T, the Er moments do not recover a fully collinear structure because
[110] is not a principal direction for the local $g$-tensor of half the Er sites.
We apply a self-consistent calculation, in the molecular field approximation,
to try and explain this evolution quantitatively; our model takes into account
the crystal field interaction previously studied in Refs.\onlinecite{Champion03} and
\onlinecite{Cao09}, together
with an anisotropic molecular field tensor, arising from exchange couplings
with the 6 nearest neighbours of a given ion. We show that we can account for the
field evolution of the magnetic structure in the AF phase (except in a limited field
range below $H_c$) by the same molecular field tensor which accounts for
the thermal variation of the local susceptibility in the paramagnetic phase \cite{Cao09}.
We also apply our model to compute the lowest excitation energies at 0.05\,K, the heat
capacity in the paramagnetic phase, and the ($H,T_{\rm N}$) phase diagram; a reasonable agreement is obtained with the experimental data of Refs.\onlinecite{Champion03} and \onlinecite{Ruff08}.
\begin{figure}
\resizebox{0.45\textwidth}{!}
{\includegraphics{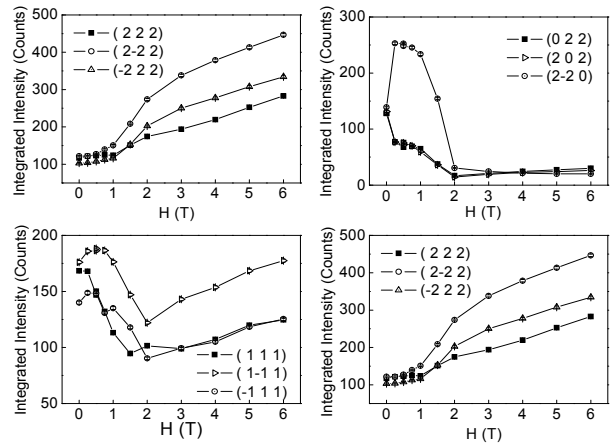}}
\caption{
Integrated intensities versus the field $H$ applied along
[110] of some typical Bragg peaks measured at 0.3\,K in \erti.}
\label{bragg}
\end{figure}

\section{Experimental}
A single  crystal of \erti\ was grown
by the floating-zone technique, using a mirror furnace. It was placed at the
bottom of a dilution inset, inside a superconducting coil. The vertical axis
of the magnetic field was aligned with the [110] axis, with a slight
misorientation discussed below.
Neutron diffraction measurements were performed at the ORPH\'EE
reactor of the Laboratoire L\'eon Brillouin, on the Super-6T2 spectrometer
in the unpolarized neutron version, with an incident neutron wavelength
$\lambda_n$=0.9\,\AA. The nuclear structure was characterized by zero field
neutron diffraction at two temperatures above $T_{\rm N}$,
 100\,K and 5\,K, allowing
the lattice constant, positional parameters, occupancy factors, isotropic
temperature factors and extinction parameters to be refined, within
the space group {\it Fd$\bar{3}$m}. To determine the magnetic structures,
about 300 Bragg peaks were collected at 0.3\,K, in zero field and for 10 field
values in the range 0 - 6\,T.

\section{Evolution of the magnetic structure}
Figure \ref{bragg} shows the field dependence of the integrated intensities
of some particular Bragg peaks. The ($h,-h,l$) peaks, situated in the
horizontal plane, have a field
dependence similar to that previously measured \cite{Ruff08}. The lifting counter geometry of the diffractometer allowed us to measure the equivalent out-of-plane ($h,h,l$) reflections plotted in the figure,
 which show different field dependencies.
These allow several regions of interest to be distinguished. The low field
region $H<0.1$\,T shows opposite
variations of the (2$\bar 2$0) and (022)/(202) Bragg peaks, which are clearly
assigned to the reorientation of the magnetic domains by the applied field.
At higher fields, a single domain is stabilized, and the variations of the
Bragg peaks arise from moment reorientations inside a tetrahedron.
The (002) and (111) Bragg peaks show an extremum at 1.5\,T, somehow below the
critical field. The evolution of the magnetic peaks with the field is well accounted
for by the magnetic refinements described below.
\begin{figure}
\resizebox{0.42\textwidth}{!}
{\includegraphics{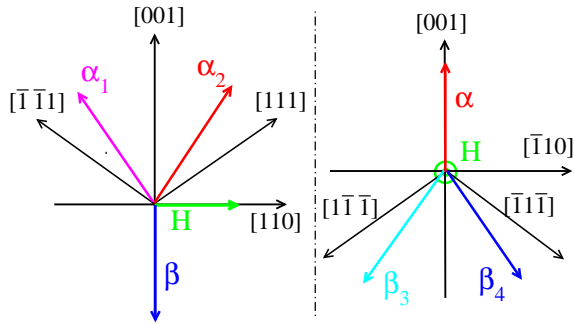}}
\caption{[Color on line] Zero-field configuration in the $\psi_2$ state in
the ($\bar 1$10) (left) and (110) (right) plane. The [110]
incipient field direction is indicated by a green arrow.
The local ternary axis (hard magnetic axis) for the $\alpha_1$ (resp.
$\alpha_2$) moments is [111] (resp. [$\bar 1\bar 1$1]); for the $\beta_3$
(resp. $\beta_4$) moments, it is [$\bar 1 1 \bar 1$] (resp. [1$\bar 1\bar 1$]).
 This configuration corresponds to that of the single domain occurring above 0.1\,T.
The upper (resp. lower) hemisphere referred to in the text is that containing the
positive (resp. negative) part of the [001] axis.
\label{strh0}}
\end{figure}

Magnetic refinements were performed with the program Fullprof
\cite{Juan_fullprof}. In zero field, the magnetic structure is a {\bf k}=0
structure,  which means that the 4 tetrahedra in the cubic unit cell have the
same moment orientations. These orientations are those corresponding to
the $\psi_2$ state, with moments in the easy planes along $<$211$>$
axes. The magnetic moment at 0.3\,K was refined as $m$=3.25(20)\,\mub\ per
Er ion. The domain populations were refined, yielding three equally
populated magnetic domains (together with the opposite domains giving the same
contribution), as also found by polarized neutron measurements \cite{Poole07}.

Above 0.1\,T, good refinements (with typical agreement factors R$_F$=5\%) were
obtained by considering a single magnetic
domain. The evolution of the magnetic structure with the field was determined
by refining the integrated magnetic intensities, with moment values and
angles as parameters. For the [110] field direction, the Er sites
split into two sets, or chains: the $\alpha$ sites, with a ternary axis at an
angle $\theta = \arcsin (1/\sqrt{3}) \simeq 35.3^\circ$ from {\bf H}, and the
$\beta$ sites with their ternary axis perpendicular to the field.
In the $\psi_2$ state, the $\alpha$ sites split further into $\alpha_1$ and
$\alpha_2$, with moments which are not symmetrical with respect to the field
direction. We also split here the $\beta$ moments into $\beta_3$ and $\beta_4$
(see Fig.\ref{strh0}). In our experimental setup, there is a slight
misorientation of the applied field with respect to the [110] direction:
the polar angle $\theta$ of the field is 93.8$^\circ$ (instead of 90$^\circ$)
and its azimuthal angle $\varphi$ is 45.8$^\circ$ (instead of 45$^\circ$). In
the calculation described in the following section, we
will neglect the small azimuthal misorientation and assume
$\varphi = 45^\circ$, but we take into account the exact $\theta$ value.
In the refinements, in order to reduce the number of fitting parameters, we
assumed equal magnitudes for the two $\beta$ moments, which was {\it a
posteriori} justified by the calculation. The preferred domain orientation
(above a field of about
0.1\,T) is such that the $\alpha$ moments lie in the upper hemisphere.
\begin{figure}
\resizebox{0.42\textwidth}{!}
{\includegraphics{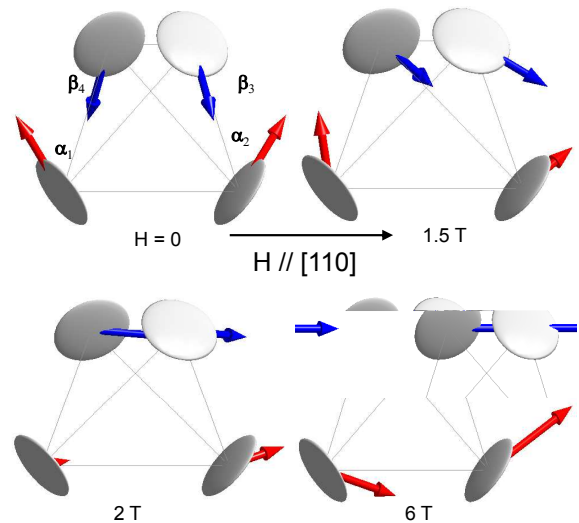}}
\caption{[Color online] Evolution of the spin structure inside a tetrahedron
for selected fields. The flat disks visualize the easy (111) planes. The
$\alpha$ moments are shown by red arrows and the $\beta$ moments by blue arrows.
with lengths proportional to the moment values.}
\label{fig3}
\end{figure}

\begin{figure*}
\resizebox{0.8\textwidth}{!}
{\includegraphics{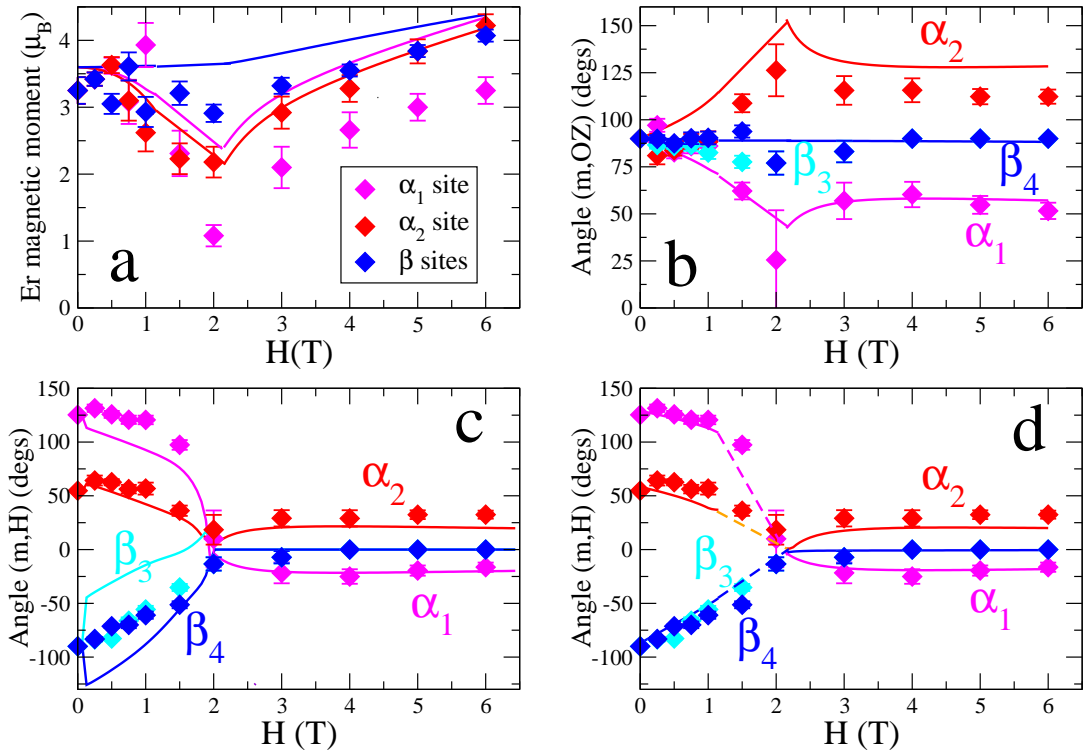}}
\caption{[Color on line] Variation with the field, applied along [110], of the
moment magnitude ({\bf a}) and orientation
with respect to the local ternary axis ({\bf b}) and to the field direction
({\bf c} and {\bf d}), for the 4 Er sites in Er$_2$Ti$_2$O$_7$ at 0.3\,K.
Our convention for the angle with the field direction is that it is negative
if the moment lies in the lower hemisphere.
Solid lines in ({\bf a})-({\bf c}) are self-consistent calculations with $\lambda_\perp =-0.51$\,T/$\mu_B$  and $\lambda_\Vert = -0.06$\,T/$\mu_B$; Solid lines in ({\bf d}) are calculations with $\lambda_X = -0.54$\,T/$\mu_B$,
$\lambda_Y = -0.48$\,T/$\mu_B$ and $\lambda_\Vert = -0.06$\,T/$\mu_B$; dashed lines are interpolated ones.
\label{mhvar}}
\end{figure*}

The moment orientations in a given tetrahedron are sketched for selected
fields in Fig.\ref{fig3}. All the moments are seen
to rotate towards the field direction (see {\bf c} or {\bf d} in Fig.\ref{mhvar})
up to a field of about 2\,T, which can be identified with the quantum critical field
$H_c$ of Ref.\onlinecite{Ruff08}. For this field value, all the moments are practically
aligned along the field, so that $H_c$ can be considered as the spin-flip field of the
\erti\ magnetic structure for the [110] field direction.
On further increase of $H$ beyond 2\,T, the $\beta$ moments remain along the
field whereas the $\alpha$ moments tend towards an asymptotic orientation, symmetric
with respect to {\bf H}.

The data points in Fig.\ref{mhvar} represent the evolution of the moment values and
orientations at the 4 Er sites as the field increases. Pronounced anomalies are seen
in the quantum critical region. As a major effect, all moment values show a minimum
at $H_c$=2\,T (Fig.\ref{mhvar} {\bf a}). This minimum is much more pronounced for the
$\alpha$ moments, which get the closest to their hard local ternary axis for this field
value. The $\alpha_1$ moment decreases down to 1\,\mub\, whereas the $\alpha_2$ moment
reaches 2\,\mub. As to the $\beta$ moments, which remain in their easy plane while
rotating, they show a much shallower minimum of about 3\,\mub.

The evolution of the angles of the moments with the field is shown in Figs.\ref{mhvar}
{\bf c} and {\bf d}. At 2\,T, all the angles are close to zero. The ``flip'' of the $\alpha_1$
moment is reflected by the change of sign of its angle with {\bf H}, when it enters the lower
hemisphere (see Fig.\ref{strh0}) above 2\,T. By contrast, the $\alpha_2$ moment approaches
the field direction at 2\,T, then tilts away, always remaining in the upper hemisphere.
Above 2\,T, the two $\alpha$ moments tend towards an orientation at 20-25$^\circ$ on either
side of the field. The $\beta$ moments show a much smoother field variation: they progressively reorient towards the field and remain aligned along {\bf H}
from 2\,T upwards. The angles of the moments with their local ternary axis (referred to as
OZ in Fig.\ref{mhvar} {\bf b}) confirm that the $\beta$ moments always remain in their easy
plane, while the $\alpha$ moments approach their hard axis in the critical region.

Our model calculation, to be described in the next section, accounts for most features of this evolution.

\section{Model calculation}
We implemented a model intended to reproduce the experimental data concerning
both the thermal variation of the local magnetic susceptibility in the
paramagnetic
phase, measured in Ref.\onlinecite{Cao09}, and the evolution with field of
the magnetic structure at 0.3\,K, measured in the present work.
This model performs mean field self-consistent calculations and uses
as ingredients the crystal field parameters of Er$^{3+}$ in
Er$_2$Ti$_2$O$_7$ as in Ref.\onlinecite{Cao09}, and anisotropic two-ion
exchange of the type:
\begin{equation}
{\cal H}_{\rm ex} = - {\cal J}_\Vert S_{1Z}S_{2Z} - {\cal J}_\perp (S_{1X}S_{2X}
+S_{1Y}S_{2Y}),
\label{ech}
\end{equation}
where ${\cal J}_\Vert$ and ${\cal J}_\perp$ are the components of the
exchange tensor in the local frame with axial symmetry. We shall also
consider the effect of a slight exchange anisotropy within the easy plane.
We consider each ion in a tetrahedron to be exchange-coupled to
its 6 nearest neighbours, resulting in an anisotropic molecular field tensor
$\tilde \lambda$ such that the molecular field acting on this ion writes:
\begin{equation}
{\bf H}_{\rm mol} = \frac{1}{6}\tilde \lambda \sum_{k=1}^6 {\bf m}_k,
\label{hmol}
\end{equation}
where the sum runs over the 6 nearest neighbors. The relationship between
the components of the $\tilde \lambda$ and $\tilde {\cal J}$ tensors is:
\begin{equation}
\lambda_i = 6 {\cal J}_i\ (\frac{g_{\rm J}-1}{g_{\rm J}})^2\ \frac{1}{\mu_{\rm B}^2}.
\label{tens}
\end{equation}
A self-consistent treatment involving the 4 Er moments, each with
its 3 components, is performed, the only parameters being thus the 2
components of the $\tilde \lambda$-tensor.
The dipolar coupling is not included in the calculation.
\begin{figure}
\resizebox{0.42\textwidth}{!}
{\includegraphics{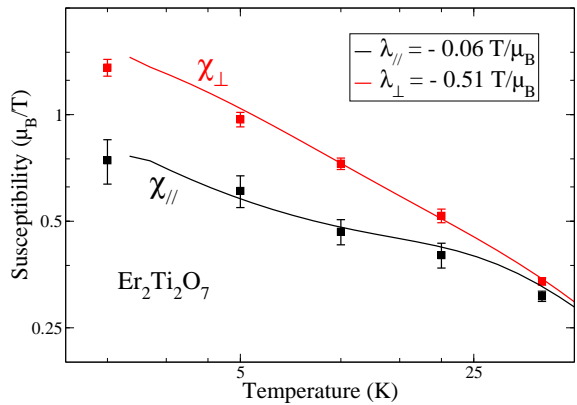}}
\caption{[Color on line] Low temperature region of the thermal variation of
the longitudinal and transverse susceptibilities
in Er$_2$Ti$_2$O$_7$. The data are taken from Ref.\onlinecite{Cao09} and the lines are
calculated using the self-consistent model described in the text.}
\label{xiT}
\end{figure}

We apply first this model to the paramagnetic susceptibility and to the
zero-field AF phase. In the paramagnetic phase, the $\chi_\Vert(T)$ and $\chi_\perp(T)$
data measured in Ref.\onlinecite{Cao09} are very well reproduced using the above
model with the axially symmetric AF molecular field tensor:
$\lambda_\Vert = -0.06(3)$\,T/\mub\ and $\lambda_\perp = -0.51(4)$\,T/\mub\
(see Fig.\ref{xiT}). These values are close to those obtained in Ref.\onlinecite{Cao09}
using a simpler single sublattice model ($\lambda_\Vert=-0.15$, $\lambda_\perp=-0.45$
\,T/\mub). The planar anisotropy of the $\tilde \lambda$ tensor is strong
($\lambda_\perp/\lambda_\Vert \simeq$10) and reinforces that of the crystal field.
In the AF phase in zero-field, the 4 Er moments in the $\psi_2$ ground state
comply with the rule: $\sum_j {\bf m}_j$=0, whence, for instance,
${\bf m}_2+{\bf m}_3+{\bf m}_4=-{\bf m}_1$. Therefore, the molecular field
acting on ion $i$ is, according to Equ.(\ref{hmol}): ${\bf H}_{\rm mol}^i = -
\frac{1}{3} \lambda_\perp {\bf m}_i$, since the spontaneous moments lie in
the easy plane. According to mean field theory, the transition temperature is given by:
$k_{\rm B}T_{\rm N} = m_0 H_{\rm mol}$, hence in the $\psi_2$ state of \erti:
$k_{\rm B}T_{\rm N} = \frac{1}{3} \vert \lambda_\perp \vert m_0^2$, where the
saturated moment $m_0$ and $\lambda_\perp$ must be obtained self-consistently in
zero applied field and near 0\,K. In order to obtain a transition temperature of
1.2\,K, one needs the value $\lambda_\perp = -0.435$\,T/\mub, which is remarkably close
to 
 that derived in the paramagnetic phase. This coherence gives
confidence in the applicability of the molecular field approximation
in the paramagnetic and AF phases of \erti. The associated
transverse exchange integral is ${\cal J}_\perp=-1.75$\,K, and the $T=0$ molecular
field $H_{\rm mol}=0.5$\,T. The zero-field spontaneous moment is $m_0=3.52$\,\mub, close to
the experimental value 3.25(20)\,\mub, but somewhat higher than 3.00(05)\,\mub\
measured in Ref.\onlinecite{Champion03}.

We then used the self-consistent exchange and crystal field model described above
to calculate the evolution of the magnetic structure with increasing field.
In a first step, we chose the two components of the axially symmetric $\tilde \lambda$
tensor as determined from the fit of the paramagnetic susceptibility ($\lambda_\perp=-0.51$\,T/\mub\ and $\lambda_\Vert = -0.06$\,T/\mub). The results are
displayed in Figs.\ref{mhvar} {\bf a}, {\bf b} and {\bf c} as solid lines.
The calculation does not exactly match the data,
but it captures the main trends of the rotation of the Er moments as the field is
increased. Remarkably again, using the $\tilde \lambda$ tensor derived in the paramagnetic phase, the value of the critical or ``spin-flip'' field $H_c \simeq$2\,T is correctly
reproduced. This value strongly depends on $\lambda_\perp$ and very
little on $\lambda_\Vert$. For a simple AF structure with moments perpendicular to the
field, the spin-flip field in the presence of anisotropic $\tilde g$ and $\tilde \lambda$ tensors writes:
\begin{equation}
 H_{\rm sf} = \frac{1}{2} g_\Vert \mu_{\rm B} \ \vert \lambda_\Vert + (\frac{g_\perp}{g_\Vert})^2
          \lambda_\perp \vert.
\label{hcAF}
\end{equation}
Since its moment configuration is more complicated than a two-sublattice AF structure, this expression cannot be directly applied to \erti, but it accounts for the strong dependence of
$H_{\rm sf}$ on $\lambda_\perp$ since $g_\perp/g_\Vert >$1 in \erti.

\begin{figure}
\resizebox{0.44\textwidth}{!}
{\includegraphics{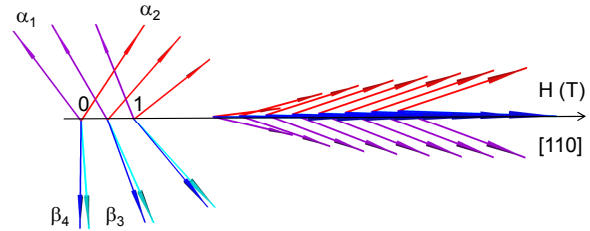}}
\caption{[Color on line] Evolution of the Er magnetic moments with the magnetic field at 0.3K, as calculated from the self-consistent calculation with with $\lambda_X = -0.54$\,T/$\mu_B$, $\lambda_Y = -0.48$\,T/$\mu_B$ and $\lambda_\Vert = -0.06$\,T/$\mu_B$. Arrows are proportional to the moment values. Each field step corresponds to 0.48~T.}
\label{fish}
\end{figure}

 However, as seen in Fig.\ref{mhvar} {\bf c}, there is a
discrepancy concerning the rotation of the $\beta$ moments: the calculation yields
different variations of the angle with the field for the two sites, whereas the
data present identical variations within experimental uncertainties. This is due to
a moment reorientation occurring in the calculation at a very small field (about 0.03\,T):
the moments jump from the $\psi_2$ state, with moments along the local OX axis of the $<$211$>$ family, to a configuration where each moment lies at about 38$^\circ$ from OX, which probably corresponds to the (small) potential well
created by the crystal field interaction. Since this reorientation is not observed in the
data, one has to devise a mechanism which would reinforce the potential well along the
local OX axis, at least at low field. For this purpose, in a second step, we lifted the in-plane degeneracy of the molecular field tensor by about 5\%, through the introduction a slightly higher (absolute) value for $\lambda_X$
than for $\lambda_Y$. The result of the calculation, with $\lambda_X = -0.54$\,T/\mub\
and $\lambda_Y=-0.48$\,T/\mub, is shown for the angle $\theta_{\rm H}$=({\bf m},{\bf H}) in
Fig.\ref{mhvar} {\bf d} (the other calculated quantities remaining practically unchanged). The agreement is better for the $\beta$ moments, i.e. the $\psi_2$ configuration evolves smoothly
with increasing field, up to a field of 1.2\,T. For $1.2<H<2$\,T, the calculation presents an anomalous moment jump, still unexplained. Therefore, in Fig.\ref{mhvar} {\bf d}, we do not consider
the calculated curves in this field interval and we have drawn dashed lines interpolated between the lower and higher field regions where the model reproduces the data reasonably well. In Fig.\ref{fish}, we show the evolutions of the four moments versus the magnetic field, calculated within this last model.

Finally, the high field configuration is
the following: the $\beta$ moments lie along the field and the $\alpha_1$ and
$\alpha_2$ moments are on either side of the field, at an angle of about 20$^\circ$.
This can be understood since the [110] direction lies within the easy plane for the $\beta$
moments, whereas it is not a principal axis for the $\alpha$ moments\cite{note1}.

It is worth emphasizing here that the data cannot be reproduced with an {\it isotropic}
molecular field constant $\lambda=-0.51$\,T/\mub. In this case, the calculation yields
moments which remain in their easy plane up to the critical field of 2\,T, and this field
is not a spin-flip field since the $\alpha$ moments do not align along its direction. The $\alpha$ moment magnitude then shows no minimum at 2\,T. Therefore, our calculation,
within its limits of validity, demonstrates the existence of an anisotropic molecular field tensor in \erti.
\begin{figure}
\resizebox{0.44\textwidth}{!}
{\includegraphics{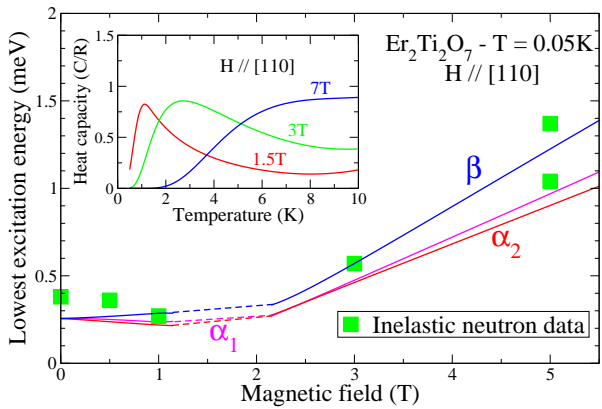}}
\caption{[Color on line] Calculated lowest excitation single ion energies for the
4 sites in Er$_2$Ti$_2$O$_7$, in the
AF ordered phase at 0.05\,K as a function of the applied magnetic field. The
square dots are the (estimated) experimental values from Ref.\onlinecite{Ruff08},
for fields where a clearly non-dispersive level is observed in the inelastic
neutron spectra. Insert: calculated thermal variation of the heat capacity
for $H$=1.5, 3 and 7\,T, obtained from the single ion energies.}
\label{zeem}
\end{figure}

\section{Discussion}
In this section, we show that the model described above also accounts relatively well for
other experimental results related to the spin excitations in \erti, as previously measured by
inelastic neutron scattering and heat capacity. In Refs.\onlinecite{Champion03} and
\onlinecite{Ruff08}, a dispersionless mode was observed at about 0.4\,meV for $H$=0, which
further splits into higher energy modes when the field increases above the critical field.
In first approximation, mostly valid in the middle of the Brillouin zone, such flat modes
may be attributed to the splitting of the ground Kramers doublet under the coupled influence
of the external and exchange fields. In this picture, the splitting into several modes
induced by the field can be understood by the different molecular fields experienced by the
$\alpha$ and $\beta$ moments. In Fig.\ref{zeem}, the excitation energies obtained by our model
are compared with the energies of the flat modes determined in
Ref.\onlinecite{Ruff08} (square dots in Fig.\ref{zeem}). Below 2\,T, the calculated energies
are almost field independent around 0.25\,meV, and above this threshold field, they increase linearly with slightly different slopes. The data
match reasonably well the calculated values, except at zero and low field, where
the energy of the non-dispersive mode (0.4\,meV) is significantly above the calculation.
One expects that our
single ion mean field energy calculation holds at high field, where exchange is small
with respect to the Zeeman energy, but fails at low field. Indeed, at low field, one should consider, for instance, the system of 4 exchange coupled Er moments on a tetrahedron and
calculate the whole spin-wave spectrum, taking altogether crystal field, exchange and Zeeman
interactions into account.

In Ref.\onlinecite{Ruff08}, the variation with field of the heat capacity in \erti\ was
attributed to a crossover  towards a quantum high field paramagnetic state reminiscent
of that observed in the quantum critical magnet LiHoF$_4$. We calculated the in-field heat capacity $C_p(T)$ in \erti\ below 10\,K,
using the crystal field and Zeeman energies of all the levels (see insert in
Fig.\ref{zeem}). Whereas this simple Schottky-type calculation cannot capture
the anomaly at the phase transition for low fields, it reproduces correctly the main
features of the high field heat capacity data of Ref.\onlinecite{Ruff08}. Especially,
the smearing of the temperature dependence of $C_p(T)$ with increasing field can be correctly
reproduced in this simple approach, which does not involves quantum fluctuations explicitly,
besides the crystal field interaction.
\begin{figure}
\resizebox{0.42\textwidth}{!}
{\includegraphics{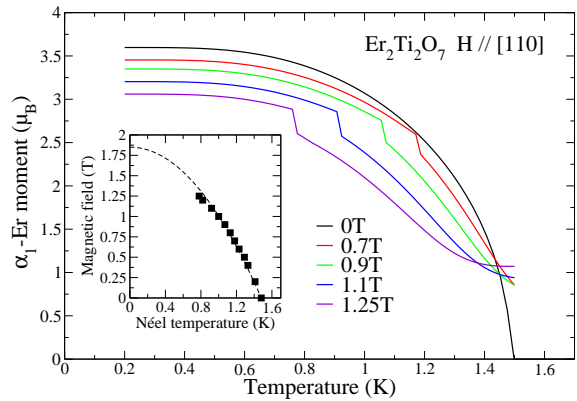}}
\caption{[Color on line] Calculated thermal variation of the $\alpha_1$ Er
moment in the presence of a magnetic field applied along [110], for different
field values below 1.3\,T, upper limit of validity of our model in the AF
phase. Insert: ($H,T_{\rm N}$) phase diagram obtained from the break-points of the $\alpha_1$-site $m(T)$ curves. The dashed line represents the law
$H=H_c [1-(\frac{T_{\rm N}}{T_{\rm N}^0})^2]$, with $H_c$=1.85\,T and
$T_{\rm N}^0$=1.48\,K.}
\label{hcrit}
\end{figure}

Finally, we also applied our model to the calculation of the ordering temperature
as a function of the applied field, i.e. to the ($H,T_{\rm N}$) phase diagram.
With the  $\tilde \lambda$-tensor determined above ($\lambda_x=-0.54$, $\lambda_y=-0.48$
and $\lambda_\Vert=-0.06$\,T/\mub), we calculated the thermal
variation of the Er moments for different field values in the temperature
range 0.2 - 2\,K, in the field range below 1.3\,T, where the model yields good
agreement with experiment. As can be seen in
Fig.\ref{hcrit}, which displays the thermal variation of the $\alpha_1$
moment magnitude, the ordering temperature, defined as the temperature where
the moment variation shows a jump, decreases with increasing field \cite{note}.
This reflects the competition between the Zeeman and exchange
energies, which leads to the quantum critical point when the Zeeman coupling
overwhelms exchange and prevents a spontaneous moment configuration to set in.
The low field part of the ($H,T_{\rm N}$) phase diagram is shown in the insert of
Fig.\ref{hcrit}. Our calculation is in qualitative agreement with the
diagram obtained from the heat capacity data of Ref.\onlinecite{Ruff08}. The curve
$H=f(T_{\rm N})$ is seen
to follow a power law (see dashed line in the insert of Fig.\ref{hcrit}), with an
exponent $n=2$. Extrapolation of this law to zero yields a critical field
$H_c=f(T_{\rm N}=0)$=1.85\,T, close to the experimental value.

\section{Conclusion}
Using neutron diffraction in the AF phase of the planar pyrochlore \erti\ with a
magnetic field applied along [110], we performed a quantitative study of the field
evolution of the magnetic structure throughout the quantum critical point at
$H_c \simeq$2\,T. For this critical field value, the AF magnetic structure has ``flipped'',
i.e. all the Er moments are aligned along the field and their magnitude reaches a
minimum. The strong decrease of the moment magnitude at $H_c$ reflects the spin
wave damping and enhanced quasi-elastic scattering previously observed in neutron
experiments. A four sublattice self-consistent calculation, taking into account exchange,
Zeeman and crystal electric field interactions, accounts for most of the characteristics
of the field induced magnetic structure, from zero field up to well above $H_c$. It also
explains semi-quantitatively the field dependence of the heat capacity, of the dispersionless
inelastic mode and of the transition temperature. The comparison between model and experiment
brings out the strongly anisotropic exchange interaction as a necessary ingredient to explain
all the features of the Quantum Critical Point. The critical field can then be defined as the
field value for which the Zeeman energy overcomes the exchange energy, taking the anisotropic
exchange tensor into account. The anisotropic exchange, already outlined by local
susceptibility measurements in the paramagnetic phase, should also influence the spin-wave
excitation spectrum in the AF ordered phase.

\begin{acknowledgments}
We acknowledge very fruitful discussions with C. Lacroix (Institut Louis N\'eel, Grenoble)
and J. Robert (LLB Saclay). Huibo Cao acknowledges supports from the Triangle de la Physique during his post-doctoral training.
\end{acknowledgments}


\begin{thebibliography}{99}
\bibitem{note1}
Actually, for an isolated  Kramers doublet, if the field is at an angle $\theta$ from the symmetry axis,
then the angle $\gamma$ of the magnetic moment with the axis is given by: $\tan \gamma =
(\frac{g_\perp}{g_\Vert})^2 \tan \theta$. Inserting the zero-field g-values in this expression
leads to an asymptotic angle of the moment with respect to {\bf H} of $\simeq 40^\circ$,
instead of the actual 20$^\circ$. This discrepancy
arises from the mixture of the excited crystal field states into the ground state by the Zeeman
effect, which modifies the g-values.
\bibitem{note}
With the mean value $\lambda_\perp=-0.51$\,T/\mub, the calculated $T_{\rm N}$ value in zero
field is 1.48\,K, rather than the experimental value 1.2\,K. This probably reflects a small
imperfection of the model, which requires a 15\% higher $\lambda_\perp$ value to account for
the in-field properties than for the zero-field $T_{\rm N}$ value.
\end{thebibliography}
\end{document}